\documentclass{elsart}
\begin{document}

\def\be{\begin{equation}}
\def\ee{\end{equation}}
\def\beq{\begin{eqnarray}}
\def\eeq{\end{eqnarray}}
\def\s{\sigma}
\def\G{\Gamma}
\def\F{_2F_1}
\def\an{analytic}
\def\ac{\an{} continuation}
\def\hsr{hypergeometric series representations}
\def\hf{hypergeometric function}
\def\ph{photon}
\def\yy{photon-photon scattering}
\def\ndim{NDIM}

\begin{frontmatter}

\title{\bf NEGATIVE DIMENSIONAL INTEGRATION FOR MASSIVE FOUR POINT
FUNCTIONS--II: NEW SOLUTIONS}
\author{\underline {Alfredo T. Suzuki} and Alexandre G. M. Schmidt}
\footnote{E-mail:suzuki@power.ift.unesp.br, schmidt@power.ift.unesp.br}
\address{Instituto de F\'{\i}sica Te\'orica -- Universidade Estadual
Paulista, R.Pamplona, 145, S\~ao Paulo SP, CEP 01405-900, Brazil}
\begin{abstract}
In this sequel calculation of the one-loop Feynman integral pertaining
to a massive box diagram contributing to the photon-photon scattering
amplitude in quantum electrodynamics, we present the six solutions as
yet unknown in the literature. These {\bf six new solutions} arise
quite naturally in the context of negative dimensional integration
approach, revealing a promising technique to handle Feynman
integrals. 
\end{abstract} 

\end{frontmatter}
\newpage

\section{Introduction.}

In our previous paper\cite{box} we calculated the Feynman integral
pertaining
to the photon-photon scattering amplitude in Quantum
Electrodynamics (QED)
using a technique we call negative dimensional
integration method
(\ndim{})\cite{halliday,halliday2,suzuki2,suzuki1}.
One of the outstanding
features of \ndim{} is that the complexities of
performing $D$-dimensional
momentum integrals are transferred to the
easier task of solving systems of
linear algebraic equations. In
\cite{box} \ndim{} has allowed us to recover
very easily the two known
\hsr{} for the pertinent Feynman integral. Another
outstanding feature
of \ndim{} --- and this is in our opinion its greatest
potential ---
is that it gives simultaneously six new results for the integral
in
question in a very straightforward manner. The aim of the present paper
is
to consider these new results. Each of them is valid in certain
regions of
external momenta and are related to the others by \ac{},
either directly or
indirectly. 

Compared to the traditional methods
of working out Feynman integrals, \ndim{} is by far simpler and more
straightforward. The difficulties of calculating
parametric integrals
--- if one is handling a Feynman integral in the
standard
way\cite{zuber} --- or solving contour integrals ---
if one is using the
Mellin-Barnes' integral representation for massive
propagators\cite{boos,stand,jmp} ---
are with finesse embedded in an
easier problem of solving systems of linear
algebraic
equations.

The Mellin-Barnes' integral technique can provide us with
only two types of
results: either the resulting function will depend
on, say, $p^2/m^2$, or its
inverse, $m^2/p^2$, depending on whether
one chooses to close the contour to
the left or to the right in the
complex plane. The resulting functions are in
general hypergeometric
ones. However, if we have more than one external
momentum (as in our
present case) and/or internal mass, there are several
other
combinations of dimensionless variables that define distinct
regions of
external momenta. It is then clear that the Mellin-Barnes'
technique will be
limited to give simultaneously only two among those
several possibilities. Not
so with \ndim{}, where many of the
possibilities are accounted for at the same
time. Of course, in both
cases, one can always construct other power series
representations by
means of suitable analytic continuations as long as the
formula for
such extensions be known. \ndim{} has then clear advantages over
other
techniques in this sense.

Hypergeometric functions of one and two
variables have lots of well-known \ac{}
formulas but as the number of
variables increases --- as far as we know --- the
fewer the known
relations\cite{appel} are. On the other hand, since \ndim{}
provides
us with very many simultaneous results, which in principle must
be
connected by \ac{}, we come to the realization that it is not only
a very
powerful technique to work out Feynman integrals but an elegant
approach to
check on \ac{} properties of the resulting functions as
well. Consider our
present case: Altogether we have eigth distinct
solutions for the Feynman
integral for the \yy{}, two of which have
already been considered in our
previous paper and six new ones, which
are connected with each other by
suitable \ac{} formulas.

One could
rightfully ask: Why do we want so many distinct results at the
same
time? We give some good arguments for this. Firstly, if we have
only
one\footnote{via parametric integration} or at most
two\footnote{via
Mellin-Barnes' integration} results in distinct
regions of the external
momenta, all the other regions must be worked
out all the way through the \ac{}
formulas, which is not always an
easy task to perform and certainly is very
much time-consuming.
Secondly, the important special case of forward scattering
in the
relativistic regime cannot be dealt with if one has only the two
known
\hsr{} for the Feynman integral relative to the \yy{}. These
series are
unsuitable for handling this special case because of the
very nature of their
variables. The same reasoning applies to the
backward scattering. Thirdly, our
results are expressed in a compact
form that can be transformed --- by an
appropriate integral
representation --- into the more cumbersome standard form
in terms of
dilogarithms, if one wants to do so. Fourthly, we can identify
the
branch points and singularities of Feynman integrals directly from
their
\hsr{}. Fifthly and lastly, since any two distinct solutions are
related by
\ac{}, \ndim{} is an elegant and economical way of
obtaining \ac{} formulas
among hypergeometric series (see Section
3).
 
The outline for this paper is as follows: In Section 2 we write
down the new
results for the Feynman integral in $D$-dimensions and
arbitrary powers of
propagators much the same as in dimensional
regularization\cite{thooft}. In all
of the resulting expressions we
have singularities (poles) either when we want
to take the physical
limit of $D=4$ or in the limits of unity powers for
propagators, so
that Section 3 is devoted to analyse the special cases we
are
interested in by introducing suitable regularization parameters.
Four of the
solutions remain divergent even after the regularization
procedure and deserve
our special analysis and discussion. We also
examine the convergence regions
for the new results and draw diagrams
for all of them. Finally in the last
section, Section 4, we make our
concluding remarks and present some further
challenges and
applications for \ndim{}.

\section{New Results from NDIM.}

In our previous paper\cite{box} we studied the integral,

\begin{eqnarray}
J(i,j,k,l;m) \label{I} &=& \int \d^D\!q\;\left(q^2-m^2\right)^i
\left[(q-p)^2-m^2\right]^j\left[(q-k_1)^2-m^2\right]^k
\nonumber\\ 
&& \times\left[(q-k_2)^2-m^2\right]^l, 
\end{eqnarray}

whose counterpart in positive $D$ is 

\begin{eqnarray*}
K(i,j,k,l;m)  &=& \int \frac{\d^D\!q\;}{\left(q^2-m^2\right)^i
\left[(q-p)^2-m^2\right]^j\left[(q-k_1)^2-m^2\right]^k}
\nonumber\\ 
&& \times\frac{1}{\left[(q-k_2)^2-m^2\right]^l}, 
\end{eqnarray*}

which is relevant to the \yy{} in QED. In particular, we are interested in the
case $J(-1,-1,-1,-1;m) \equiv K(1,1,1,1;m)$.

Hereafter, unless otherwise noted, we closely follow the notation of
\cite{box}. So, let $\s=i+j+k+l+\half D$, where $D$ is the space-time dimension
and let the Pochhammer symbol be

\[ (a|k) \equiv (a)_k = \frac{\G(a+k)}{\G(a)}. \]

Here we remind ourselves that the results of (\ref{I}) come from the solutions
of a system of linear algebraic equations and that for this particular integral
the system is such that has twenty-one solutions altogether, six of which are
trivial ones and fifteen of which give double series. In \cite{box} we listed
five of them which combined appropriately among themselves led to the two
well-known \hsr{}\cite{stand} of the Feynman integral. Below we give the
remaining ten solutions for the system, which combined appropriately among
themselves, yield the six new solutions for the pertinent Feynman integral. In
the following we use $s$ and $t$ for the usual Mandelstam variables.

\begin{eqnarray}
\label{I6}
I_6 &=& f_6\;{\mathcal S}_1 \left(\alpha_6, \alpha'_6, \beta_6, \beta'_6, \theta_6;
\gamma_6, \theta'_6\left| \frac{-t}{s}, \frac{4m^2}{s} \right)\right. \\
\label{I7}
I_7 & = & I_6(i\leftrightarrow k, j\leftrightarrow l\; | \;s\leftrightarrow
t)\\
\label{n1}
I_8 & = & f_8\;{\mathcal S}_2 \left(\alpha_8, \beta_8, \gamma_8, \delta_8, \phi_8;
\rho_8, \phi'_8\left| \frac{-t}{s}, \frac {-4m^2}{t}\right)\right. \\
\label{n2}
I_9 & = & I_8(k\leftrightarrow l)\\
\label{n3}
I_{10} & = & f_{10}\;{\mathcal S}_2 \left(\alpha_{10}, \beta_{10}, \gamma_{10},
\delta_{10}, \phi_{10};
\rho_{10}, \phi'_{10}\left| \frac{-s}{t}, \frac {-4m^2}{s}\right)\right. \\
\label{n4}
I_{11} & = & I_{10}(i\leftrightarrow j)\\
\label{n5}
I_{12} & = & f_{12}\;{\mathcal S}_2 \left(\alpha_{12}, \beta_{12}, \gamma_{12},
\delta_{12}, \phi_{12};
\rho_{12}, \phi'_{12}\left| \frac{4m^2}{s}, \frac {-t}{4m^2}\right)\right. \\
\label{n6}
I_{13} & = & I_{12}(k \leftrightarrow l)\\
\label{n7}
I_{14} & = & f_{14}\;{\mathcal S}_2 \left(\alpha_{14}, \beta_{14}, \gamma_{14},
\delta_{14}, \phi_{14};
\rho_{14}, \phi'_{14}\left| \frac{4m^2}{t}, \frac {-s}{4m^2}\right)\right. \\
\label{n8}
I_{15} & = & I_{14}(i\leftrightarrow j)\;,
\end{eqnarray}

where we have defined the following two functions

\beq 
\label{S_1} {\mathcal S}_1(\alpha,\alpha',\beta,\beta',\theta;\gamma,
\theta'|z_1,z_2) &=& \sum_{\mu,\nu=0}^\infty \frac{z_1^\mu z_2^\nu}{\mu!\nu!}
\frac{(\alpha|\mu)(\alpha'|\nu)(\beta|\mu)(\beta'|\nu)}{(\gamma|\mu+\nu)}
\nonumber\\ 
&&\times \frac{(\theta|\mu+\nu)}{(\theta'|\mu+\nu)}\;, 
\eeq

and

\beq 
\label{S2} {\mathcal S}_2(\alpha,\beta,\gamma,\delta,\phi;\rho,
\phi'|z_1,z_2) &=& \sum_{\mu,\nu=0}^\infty \frac{z_1^\mu z_2^\nu}{\mu!\nu!}
\frac{(\alpha|\mu-\nu)(\beta|\mu)(\gamma|\nu)(\delta|\nu)}{(\rho|\mu)}
\nonumber\\ 
&&\times \frac{(\phi|\nu-\mu)}{(\phi'|\nu-\mu)}\;,
\eeq

with

\begin{eqnarray*}
f_6 & = & (-\pi)^{D/2}\left(\frac
s4\right)^\s\frac{(-i|\s)(-j|\s)}{(\half-\half\s|\s+\quart
D)(-\half\s|\s+\quart D)}\;,\\
f_8 & = & (-\pi)^{D/2}\;s^l\;t^{\s-l}\frac{(-k|-i-j-\half
D)(-i|i-k+l)(-j|\s-l)}{(-i-l+\s|i+l+\half D)}\;,\\
f_{10} & = & (-\pi)^{D/2}\;s^j\;t^{\s-j}\frac{(-l|-i-j-\half
D)(-i|i+k-l)(-j|\s-k)}{(-i-k+\s|i+k+\half D)}\;,\\
f_{12} & = & (-\pi)^{D/2}\;s^l\;(-m^2)^{\s-l}\frac{(-k|l)(\s+\half
D|l-2\s-\half D)}{(\s+\half D|2l-2\s)}\;,\\
f_{14} & = & (-\pi)^{D/2}\;s^j\;(-m^2)^{\s-j}\frac {(-i|j)(\s+\half
D|j-2\s-\half D)}{(\s+\half D|2j-2\s)}\;,
\end{eqnarray*}

with the following parameters for the function ${\mathcal S}_1$:

\begin{eqnarray*}
\alpha_6 & = & -k,\\
\alpha'_6 & = & \half-\half\s-\quart D,\\
\beta_6 & = & -l,\\
\beta'_6 & = & 1-\half\s-\quart D,\\
\theta_6 & = & -\s,\\
\gamma_6 & = & 1+i-\s,\\
\theta'_6 & = & 1+j-\s, 
\end{eqnarray*}

and the parameters for the function ${\mathcal S}_2$:
\[
\begin{array}{lcl}
\alpha_8  =  j+k+\half D, & \hspace{.5in} & \alpha_{10}  =  i+k+\half D,\\
\beta_8  =  -l,           & \hfill & \beta_{10}  =  -j,\\
\gamma_8 =  1-\half\s-\quart D, & \hfill & \gamma_{10} = 1-\half \s-\quart D,\\ 
\delta_8 = \half -\half \s - \quart D, & \hfill & \delta_{10} = \half -\half \s
- \quart D,\\
\phi_8 = -i-j-k-\half D, & \hfill & \phi_{10} = -i-k-l-\half D,\\
\rho_8 = 1+k-l, & \hfill & \rho_{10} =  1+i-j,\\
\phi'_8 = 1-i-k-\half D, & \hfill & \phi'_{10} = 1-i-l-\half D\;, 
\end{array}
\] 
\[
\begin{array}{lcl}
\alpha_{12} = \half +\half i+\half j +\half k-\half l, & \hspace{.5in} & \alpha_{14} = \half +\half i-\half j +\half k+\half l ,\\
\beta_{12} = -l, & \hfill & \beta_{14} = -j,\\
\gamma_{12} = -i, & \hfill & \gamma_{14} = -k,\\
\delta_{12} = -j, & \hfill & \delta_{14} = -l,\\
\phi_{12} = -i-j-k-\half D, & \hfill & \phi_{14} = -i-k-l-\half D,\\
\rho_{12} = 1+k-l, & \hfill & \rho_{14} = 1+i-j,\\
\phi'_{12} = -\half i -\half j-\half k +\half l, & \hfill & \phi'_{14} = -\half i +\half j-\half k -\half l\;. 
\end{array}
\]

Observe that when the parameters $\theta$ and $\theta'$ in ${\mathcal S}_1$ are
equal, then our defined function ${\mathcal S}_1$ becomes the known Appel's
\hf{} $F_3$, whereas when the parameters $\phi$ and $\phi'$ in ${\mathcal S}_2$ are
equal, our defined function ${\mathcal S}_2$ reduces to the known Appel's \hf{}
$H_2$ (see Section 3).

Looking carefully at these one can verify without difficulty that there are
symmetry relations among them. For example, if we make the substitution $s
\leftrightarrow t$, $i \leftrightarrow k$ and $j \leftrightarrow l$ in
(\ref{I6}) we obtain (\ref{I7}). In a similar manner, in (\ref{n1}) the
substitution $k \leftrightarrow l$ transforms it in (\ref{n2}) and the
substitution $s \leftrightarrow t$, $j \leftrightarrow k$, $i \leftrightarrow
l$ yields (\ref{n2}) $\leftrightarrow $ (\ref{n3}). There are several other
symmetry properties of the box diagram which transform one solution into
another.

Now we must combine them in such a way to have sums of linearly independent
solutions bearing the same functional variable. This is the constructive
prescription\cite{box}. We then get from the above list six types of functional
variables, that is,  six new such combinations or six new results for the
Feynman integral (\ref{I}), namely,

\be 
J_3 = I_6,\qquad\qquad J_4 = I_7, 
\ee

\be 
J_5 = I_8+I_9,\qquad\qquad J_6 = I_{10}+I_{11},
\ee 

\be 
J_7=I_{12}+I_{13},\qquad\qquad J_8=I_{14}+I_{15}. 
\ee

Note that the relevant Feynman integral is obtained via

\[
K(i,j,k,l;m) \equiv J(-i,-j,-k,-l;m)
\]

\section{Regularization and Discussion.}

Here we constrain ourselves to the special case where the integral (\ref{I}) is
the one for QED \yy{} at the one-loop level, that is, we are interested in
taking the particular values $i=j=k=l=-1$. However, these expressions become
singular when we take the referred limit and/or let $D=4$. Therefore, some kind of
regularization procedure is called for.

For the first two, i.e., $J_3$ and $J_4$, we can adopt the standard procedure
of dimensional regularization\cite{zuber}. Introduce $D=4-\varepsilon$
and expand the whole expression around $\varepsilon=0$ to get

\be \label{I6a}I_6^R = \frac{8\pi^2}{s^2}\!\!\left[\frac{-2}{\varepsilon}+
\log{(-2\pi s)} + \gamma_E\right]\!\!F_3(1,\half+\half\varepsilon,1,1+
\half\varepsilon;2+\half\varepsilon|x,y), \ee
where $x=-t/s$, $\;y=4m^2/s$, $F_3$ is a \hf{} of two variables which
is absolutely convergent for $|x|<1$ and $|y|<1$,  and $\gamma_E$ is
the Euler's constant\cite{appel,bateman}. We can write a simpler
expression by using a reduction formula \cite{appel,bateman,edinburgo},

\be \label{F3F1} F_3(\alpha,\alpha',\beta,\gamma-\beta;\gamma|x,y) = 
\frac{1}{(1-y)^{\alpha'}}F_1(\beta,\alpha,\alpha';\gamma|x,z), \ee
where $z=y/(y-1)$ and $F_1$ is another \hf{} of two variables which is
absolutely convergent in the same region of the $F_3$ above. This
function has a simple integral representation\cite{bateman},

\be \label{F1} F_1(\alpha,\beta,\beta';\gamma|z_1,z_2) =
\frac{\G(\gamma)}{\G(\alpha)\G(\gamma-\alpha)}\int_0^1
\d u\frac{u^{\alpha-1}(1-u)^{\gamma-\alpha-1}} {(1-uz_1)^\beta
(1-uz_2)^{\beta'}},\ee 
where the parameters must satisfy Re$(\alpha)>0$ and
Re$(\gamma-\alpha)>0$. It is straightforward to evaluate this integral
when the parameters take the values we have in hands.
Substituting (\ref{F1}) and (\ref{F3F1}) in (\ref{I6a}) and expanding
the \hf{} in Taylor series, we get

\beq J_3(-1,-1,-1,-1;m) &\equiv & I_6^R=\frac{8\pi^2}{s(s-4m^2)}\left[\frac{-2}{\varepsilon}-\partial_{\beta'}-\partial_{\gamma}+\log{(-2\pi s)}\right. \nonumber\\ 
& & +\left. \gamma_E +\log{\left(1-\frac{4m^2}{s}\right)}\right]F_1(\alpha,\beta,\beta';\gamma|x,z) .
\eeq

Note that there is a simple pole which we did not expect by naive
power counting. We will discuss this singularity and the one that
appears in the following solution in the next subsection. Here we
introduce the parametric derivatives\cite{stand,davyd},   

\be
\frac {\partial (\alpha|z)}{\partial\alpha}\equiv \partial_\alpha (\alpha|z) = (\alpha|z)\left[\psi(\alpha+z)-
\psi(\alpha)\right] ,\ee 
where the $\psi-$function is the logarithmic derivative of the gamma
function \cite{bateman,lebedev}. First carry out the parametric
derivatives in (\ref{F1}) then substitute the values of the parameters
and integrate. For the other terms the integral results in,

\be F_1(1,1,\half;2|x,z) = -\frac{s}{t}\frac{1}{R_{st}} 
\log{\left(\frac{1+R_{st}}{1-R_{st}} \frac{R_s-R_{st}} {R_s+R_{st}}
\right)},\ee 
where 
\be \label{R} R_s = \sqrt{1-\frac{4m^2}{s}},\qquad\quad 
R_{st} = \sqrt{1-\frac{4m^2}{t}-\frac{4m^2}{s}} .\ee

See that the limit $t\rightarrow 0$ is well-defined.

We can write down immediately the result for the integral $I_7$ by
noting that it can be transformed into $I_6$ if we make the changes
$i\leftrightarrow k$, $j\leftrightarrow l$ and $s\leftrightarrow t$, 

\beq J_4(-1,-1,-1,-1;m) &\equiv & I_7^R =
\frac{8\pi^2}{t(t-4m^2)}\left[\frac{-2}{\varepsilon}
-\partial_{\beta'}-\partial_\gamma+\log{(-2\pi t)}\right. \nonumber\\   
&& \left. +\gamma_E+\log{\left(1-\frac{4m^2}{t}\right)}\right]
F_1(\alpha,\beta,\beta';\gamma |w,w').\eeq
where $w=-s/t$ and $w'=4m^2/(4m^2-t)$. For the region of convergence
see figure 1. 

For the remaining solutions dimensional regularization
is unsuitable to
regularize their divergences. Consider for example
(\ref{n1}) where there is a
factor $(-i|i-k+l)$ which is divergent in
the particular limit we are interested
in, i.e., $i=j=k=l=-1$. This
factor has no $D$-dependence and dimensional
regularization here is
useless. What we must do is to use a different
procedure, namely,
regularizing the exponent of some of
the propagators\cite{box,stand,letb}. 

Let us then consider the
fifth solution of the Feynman integral,
$J_5$. We must regularize one
exponent of one of the
propagators, say, $k=1-\zeta$ (we could also
take the exponent $l$.).
The important point is that the final result
will be independent of this
choice. The other exponents are set to
minus one while the dimension of the
space-time remains arbitrary.
Doing this we have 

\beq \label{n1r} I_8^{R} = &(-\pi)^{D/2}&
\frac{1}{st^{3-D/2}} 
\frac{\G(3-\half D+\zeta)  \G(\zeta)\G^2(\half D-2-\zeta)}
{t^\zeta\G(1+\zeta)\G(D-4-\zeta)}\nonumber\\
&& \times H_2\left(-\zeta,\; 1, 1+\half\zeta,\;
\half+\half\zeta,1-\zeta\left|\frac{-t}{s},\frac{-4m^2}{t}\right)\right.
,\eeq 
and 
\beq \label{n2r} I_9^{R} = &(-\pi)^{D/2}&\frac{\G(3-\half D)\G^2(\half
D-2)} {st^{3-D/2}}\frac{\G(-\zeta)}{s^\zeta\G(D-4-\zeta)}\nonumber\\ 
&& \times H_2\left(0,\; 1+\zeta, 1+\half\zeta,\;
\half+\half\zeta,1+\zeta\left|\frac{-t}{s},\frac{-4m^2}{t}\right)\right.
.\eeq 

The \hf{} $H_2$ is defined by the double sum\cite{bateman}, 

\be \label{h2} H_2(\alpha,\beta,\gamma,\delta;\rho|x,y) =
\sum_{m,n=0} ^\infty \frac{(\alpha|m-n)(\beta|m)(\gamma|n)
(\delta|n)}{(\rho|m)} \frac{x^my^n}{m!n!} .\ee

The region of absolute convergence of this function $H_2(\ldots|x,y)$
is bounded by the
lines\cite{bateman},
\be
\label{regiao}
|y|<\frac{1}{1+|x|},\qquad
|x|<1,\qquad |y|<1\;,
\ee
see figure 2.

In proceeding our analysis
of the new results, for the solution $J_5$, let us
now expand the
$H_2$ function in Taylor series around $\zeta=0$, keeping terms
up to
the first order in $\zeta$
\begin{eqnarray} J_5(-1,&-1&,-1,-1;m) =
\frac{2^{5-D}(-\pi)^{D/2}\sqrt{\pi}\G(3-\half
D)\G(\half D-2)}{s
t^{3-D/2} \G(\half D-\threehalf)}\left[-\gamma_E
\right.\nonumber\\ 
&& \left. +\log{\left(\frac{s}{t}\right)}-2\psi\left(\half D-1\right)+
\psi\left(3-\half D\right)
+\frac{4}{D-4} -\partial_\alpha -
\partial_\rho\right]\nonumber\\ 
&&\times
H_2\left(\alpha,1,1,\half;\rho\left|\frac{-t}{s},
\frac{-4m^2}{t}\right)\right.,
\end{eqnarray}
where the parametric derivatives must be taken at the
point $\alpha=0;\;\;\rho=1$. The nature and meaning of these
singularities will be the touched on in the following subsection. 

In a similar manner we regularize the sixth solution.
But now we take
$i=-1-\zeta$. As a result we get, 
\beq J_6(-1,&-1&,-1,-1;m) =
\frac{2^{5-D}(-\pi)^{D/2}\sqrt{\pi}\G(3-\half D)\G(\half D-2)}{t
s^{3-D/2} 
\G(\half D-\threehalf)}\left[-\gamma_E\right. \nonumber\\ 
&& \left. + \log{\left(\frac{t}{s}\right)} -2\psi\left(\half D-1\right)+
\psi\left(3-\half D\right)
+\frac{4}{D-4} -\partial_\alpha -
\partial_\rho\right]\nonumber\\ 
&&\times H_2\left(\alpha,1,1,\half;\rho\left|\frac{-s}{t},\frac{-4m^2}{s}
\right)\right., \eeq 
and as we shall see later on, these two
functions $H_2$ are related to the
functions $F_3$ that are divergent
too. The region of convergence can be constructed as we did
above for
the $H_2$ function (see fig.3).

Consider now the seventh solution of
the Feynman integral, $J_7$. Like the
preceding case, it has a simple
pole in the exponents, so that it demands only
one suitable parameter
to regularize it. Looking at (\ref{n5}) and (\ref{n6})
we note that a
good choice to introduce our regularization parameter is to
take
$l=-1-\zeta$, while the other exponents can be set to $i=j=k=-1$
without any
problem.
Then,

\beq
\label{n5r}
I_{12}^R
&=&\frac{-\pi^2}{m^2s}\left(-\frac{1}{\zeta}-1+\log{s}+O(\zeta)\right)\nonumber\\
&&\times
{\mathcal S}_2\left(-\half+\half \zeta, 1+\zeta,1,1,3-\half
D;1+\zeta,1-\half
\zeta\left|\frac{4m^2}{s},\frac{-t}{4m^2}\right)\right.\;,
\eeq

and

\beq
\label{n6r}
I_{13}^R
&=&\frac{-\pi^2}{m^2s}\left(\frac{1}{\zeta}-1-\gamma_E-\log{(-m^2)}+O(\zeta)\right)\nonumber\\
& & \times {\mathcal S}_2\left(-\half-\half\zeta,1,1,1,3-\half
D+\zeta;1-\zeta,1+\half\zeta\left|\frac{4m^2}{s},\frac{-t}{4m^2}\right)\right.\;.
\eeq

Now expand the factors of (\ref{n5r}), (\ref{n6r}) and the series
(\ref{S2}) around $\zeta=0$ and substitute the values $\alpha=-\half,\;\;\;
\beta=\gamma=\delta=\rho=\phi=\phi'=1$. Using the fact that
$\partial_\beta+ \partial_\rho=0$ (only because these two
parameters are equal) and an analogous relation between $\phi$ and
$\phi'$, we get, in four dimensions,  

\beq J_7(-1,-1,-1,-1;m) &=& \frac{\pi^2}{m^2s}\left[2+ \gamma_E+
\partial_\alpha+  \partial_\rho -
\log\left(\frac{-s}{m^2}\right)\right] \nonumber\\
&&\times H_2\left(\alpha,1,1,1;\rho\left|\frac{4m^2}{s},\frac{-t}{4m^2}
\right.\right) .\eeq

Note that the above result is finite and that there is no dependence on
 $\phi$
and $\phi'$, so ${\mathcal S}_2$ reduces to $H_2$. The pole cancels out and then
we can take the limit of vanishing $\zeta$.

The next two solutions follow the same procedure, 
yielding

\beq J_8(-1,-1,-1,-1;m) &=& \frac{\pi^2}{m^2t} \left[2+\gamma_E+
\partial_\alpha+\partial_\rho -
\log\left(\frac{-t}{m^2}\right)\right]\nonumber\\ 
&& \times H_2\!\left(\alpha,1,1,1;\rho\left|\frac{4m^2}{t},\frac{-s}{4m^2}
\right.\right) ,\eeq
which is also finite. The region of convergence of this solution is shown in figure 4. We do not need to 
calculate the parametric
derivatives because Davydychev already did it\cite{stand}. Using the
transformation formula between $H_2$ and $F_2$\cite{edinburgo},

\beq \label{h2f2} H_2(\alpha,\beta,\gamma,\delta;\rho|x,y) &=& 
{\mathcal A}_1\;F_2\left(\alpha+\gamma, \beta, \gamma; \rho,
1+\gamma-\delta\left|x, \frac{-1}{y}\right)\right.\nonumber\\ 
&& + {\mathcal A}_2\; F_2\left(\alpha+\delta, \beta, \delta; \rho,
1+\delta-\gamma\left|x, \frac{-1}{y}\right)\right. ,
\eeq  
where we define the coefficients
\be
{\mathcal A}_1=\frac{\G(1-\alpha)\G(\delta-\gamma)}{\G(\delta)\G(1-\alpha
-\gamma)} y^{-\gamma}, \qquad {\mathcal A}_2 = {\mathcal A}_1(\gamma
\leftrightarrow \delta)
\ee
we can identify the parametric derivatives of $H_2$ with the ones of $F_2$
calculated by Davydychev. Care must be taken with (\ref{h2f2}) because
with the particular parameters we have in hands the individual terms on the
RHS are singular, but which cancels out at the end when both terms are added
together.

\subsection{Discussion.}

As we have mentioned earlier, the set of new solutions we have obtained here
contains singular solutions that deserve a closer look. Let us examine them in
order to understand the meaning and the nature of such singularities. To begin
with, let us give some arguments to show the correctness of our results. As we
had conjectured in \cite{box}, solutions containing the $H_2$ \hf{}s did in
fact appear. 

Consider the first result we obtained in our previous work\cite{box,stand},
i.e.,

\be
J_1(-1,-1,-1,-1;m) = \frac{\pi^2}{6m^4}F_3\left(1,1,1,1;
\frac{5}{2}\left| \frac{s}{4m^2},\frac{t}{4m^2}\right)\right. .
\ee

The \hf{} $F_3$ which appears here is related to the \hf{} $H_2$ via \ac{} (see
Erd\'elyi\cite{edinburgo}),

\beq
\label{f3h2} F_3(\alpha,\alpha',\beta,\beta';\gamma|x,y)&=&
{\mathcal B}_1\;H_2\left(1+\alpha-\gamma, \alpha, \alpha',\beta';1+\alpha-\beta\left|
\frac{1}{x},-y\right)\right. \\ 
&& + {\mathcal B}_2\; H_2\left(1+\beta-\gamma,\beta,\alpha',\beta';
1+\beta-\alpha\left|\frac{1}{x},-y\right)\right.\nonumber ,
\eeq

where the two coefficients are
\be {\mathcal B}_1 = \frac{\G(\beta-\alpha)\G(\gamma)}{\G(\beta)\G(\gamma-\alpha)}
(-x)^{-\alpha}\; , \qquad {\mathcal B}_2 = {\mathcal B}_1(\alpha \leftrightarrow
\beta)\;.
\ee

So, with the help of equation (\ref{f3h2}) we rewrite $F_3$ in terms of $H_2$
without worrying very much about constant factors because they arrange
themselves properly in the process. Indeed, in this case both factors on the
RHS containing gamma functions are singular (this is a special case of \ac{}
known as the logarithmic case), but whose singularities cancel out at the end,
leaving us with a finite result as it should be. Then,

\beq J_1 &\sim& F_3\left(1,1,1,1;\frac{5}{2}\left|\frac{s}{4m^2},
\frac{t}{4m^2}\right)\right.=C_1 H_2\left(
-\frac{1}{2},1,1,1;1\left|\frac{4m^2}{s},\frac{-t}{4m^2}\right)\right.+
\nonumber\\
&& + C_2 H_2\left(-\frac{1}{2},1,1,1;1\left|\frac{4m^2}{s},
\frac{-t}{4m^2} \right)\right. ,\eeq

which clearly portrays the same $H_2$ function we have in $J_7$. Conclusion:
\ndim{} provides, even if we did not know (\ref{f3h2}) {\em a priori}, the
transformation $J_1 \rightarrow J_7$, or, in other words, the \ac{} formula
$F_3 \rightarrow H_2$. Moreover, as Erd\'erlyi\cite{edinburgo} mentioned, there
is a transformation similar to (\ref{f3h2}) for the variable $y$ in $F_3$. This
will give $J_1 \rightarrow J_8$.

In order to verify that there are branch points in the Feynman integral, we can
do the following. Consider the definition of the \hf{} $H_2$ given in
(\ref{h2}). Substituting the values of the parameters --- recall that the
derivatives of an analytic function are also analytic having the same region of
convergence --- two of them cancel out and we get
\be
H_2\left(-\half,1,1,1;1\left|x,y\right)\right.=\sum_{\mu,\nu=0}^{\infty}
\frac{(1|\nu)(1|\nu)}{(\half|\nu)}\frac{(-y)^{\nu}}{\nu!}
(-\half-\nu|\mu)\frac{x^\mu}{\mu!}\;, \ee
where we have used the identity $(a|-k)=(-1)^k/(1-a|k)$. Observe that the
series in $\mu$ is a \hf{} $_1F_0$\cite{lebedev} that can be summed. It results
in the following
\be \label{sing}
H_2\left(-\half,1,1,1;1\left|x,y\right)\right.=\sqrt{1-x}\sum_{\nu=0}^{\infty}
\frac{(1|\nu)(1|\nu)}{(\half|\nu)}\frac{[-y(1-x)]^{\nu}}{\nu!}\;,
\ee
with variables $x$ and $y$ given in either $J_7$ or $J_8$. The remaining series
in $\nu$ is a $_2F_1$ \hf{} that can be written down in terms of an elementary
function and it is straightforward to show that it has branch points, see
(\ref{surface}) below.

The same procedure can be applied to $J_3$. Using (\ref{f3h2}) for the \hf{}
$F_3$, we get

\beq J_3 &\sim& F_3\left(1,1,1,\frac{1}{2};2\left|\frac{-t}{s},
\frac{4m^2} {s}\right)\right. = C_3 H_2\left(0,1,1,
\frac{1}{2};1\left|\frac{-s}{t},\frac{-4m^2}{s}\right)\right.+
\nonumber\\ 
&&+ C_4 H_2\left(0,1,1,\frac{1}{2};1\left|\frac{-s}{t},\frac{-4m^2}{s}
\right)\right. ,\eeq 
yielding $J_3 \rightarrow J_6$. The analogous transformation for the
variable
$y$ in $F_3$ yields $J_3 \rightarrow J_5$ and so on.

Using analogous routes we used above, it is possible to express
this $H_2$
function in terms of an elementary function, this time a
square root.
Considering its definition, the canceling of the
parameters $\beta$ and $\rho$
for the specified values and summing the
series in $\nu$ we get,
\be
\label{singular}
H_2\left(0,1,1,\half;1\left|x,y\right)\right. =
\sum_{\nu=0}^{\infty}\frac
{[-y(1-x)]^\nu(\half|\nu)}{\nu!} = \frac
{1}{\sqrt {1+y(1-x)}}\;,
\ee
observe that the square root in the
denominator is equal to $R_{st}$, see eq.(\ref{R}). 
An important point to note here is
that even though the results remain
divergent, they are still
connected by an \ac{} formula. The questions that
need to be addressed
then are now: {\it What does this mean?} {\it What is the
nature of
these singularities?}

First of all, it is known\cite{eden} that a
four-point graph like the one in
the \yy{} has no leading
singularities in the physical region. Such
singularities does happen
to occur in four-point functions when the
two incoming particles enter
the same vertex and the two outgoing particles
also leave the same
vertex (see Fig. 5). Since this is not our case, we thus conclude that
the singularities we have do not occur on the physical sheet, i.e.,
they are harmless\cite{eden}.

Secondly, in analytically continuing a given function from a region
${\mathcal
R}_1$ into another region, ${\mathcal R}_2$ it is important that no
singularities
be present between the regions, otherwise the result for
the \ac{} may not be
unique. The non-uniqueness always manifest itself
whenever the singularity is
of the branch point
type\cite{feshbach}. We know that for the \yy{} process we
have a
branch-cut in $s=4m^2$ in the $s$-channel, so that in carrying
out our analytic
continuation from $J_6 \rightarrow J_3$, we are
crossing this branch-cut, and
then the singularities do arise.

This naive argumentation shows us the great possibilities of \ndim{}. It
reproduces three general --- with no restriction in the parameters
---
\ac{} relations between Appel's \hf{}s which are far from trivial
to obtain (see \cite{edinburgo}). It is clear too that the technique
allows us the
bonus by-product of pinpointing singularities of Feynman
integrals.  

Eden\cite{london} devised a technique to find out the
singularities of integral representations. In \cite{eden} Eden {\it et
al} applied it in the general box diagram and the equation of Landau's 
surface --- the surface of possible singularities of a integral
representation --- is given by a 4x4 determinant. In our case (equal
mass for the virtual matter fields and on-shell photons) the Landau's 
equation\cite{zuber,eden,todorov} is, 

\be \frac{st}{4m^6}\left(\frac{st}{4m^2}-s-t\right) = 0  ,\ee
so that there are four possible solutions,
\be \label{surface} s=0,\quad t=0,\quad s=\frac{4m^2t}{t-4m^2},\quad
t=\frac{4m^2s}{s-4m^2} .\ee 
Just here it is important to observe that the two last solutions make
the \hf{} $H_2$ in (\ref{singular}) and in (\ref{sing}) singular. {\em They are branch
points of the Feynman integral}. The first two are the so-called
pseudo-threshold \cite{zuber,eden,todorov,fairlie} --- singularities of
the
Feynman integral which occur on an unphysical sheet --- see also
that the possible singularities of (\ref{I}) are
located in the
region of convergence of the two above functions, $J_3$ and $J_4$. We
think that the poles does not cancel because of this reason, the
so-called pinch singularities. We can verify, comparing the analysis
contained in  \cite{zuber}, that the last two solutions of the
Landau's equation are in fact singularities of (\ref{I}).

\section{Conclusion.}

Using the technique known as negative dimensional integration we obtained six
new results for the massive Feynman integral for the box diagram which
contributes to the \yy in QED. These results are expressed in terms of Appel's
\hf{}s $F_3$ and $H_2$. All these new results had to be regularized. Four of
them remain divergent even after the regularization procedure, but the
tenacious singularities are harmless because they do not lie in the physical
region. The other two become finite after the regularization procedure and
these are important in treating relativistic dynamics of forward scattering.
We have shown that these new solutions are correctly related with each other by
\ac{}. To further check these results, calculation of the forward amplitude 
for the \yy{} amplitude in QED is in progress.

\begin{ack}
AGMS would like to thank Viviane Lisovski for helping with the drawing of
computer generated pictures and Reinaldo L. Cavasso F$^o$(UFPR) for getting
reference \cite{edinburgo}. AGMS acknowledges financial support from CNPq
-Conselho Nacional de Desenvolvimento Cient\'{\i}fico e Tecnol\'ogico, Brazil.
\end{ack}


\begin{thebibliography}{99}

\bibitem{box}
A.T.Suzuki, A.G.M.Schmidt, submitted
to Nucl.Phys.{\bf B} (1997).

\bibitem{halliday}
I.G.Halliday, R.M.Ricotta, Phys.Lett.{\bf B193},2 (1987)241.
R.M.Ricotta,  {\it Topics in Field Theory} (Ph.D. Thesis,Imperial
College, 1987). 

\bibitem{halliday2}
G.V.Dunne, I.G.Halliday, Phys.Lett.{\bf B193},2(1987)247.

\bibitem{suzuki2}
A.T.Suzuki, R.M.Ricotta, {\it XVI Brazilian Meeting on Particles 
and Fields}, 386(1995), C.O.Escobar(Ed.).

\bibitem{suzuki1}
A.T.Suzuki, R.M.Ricotta, {\it Topics on Theoretical Physics - 
Festschrift for P.L.Ferreira}, 219(1995), V.C.Aguilera-Navarro {\it et
al} (Ed.).

\bibitem{zuber}
C.Itzykson, J-B.Zuber, {\it Quantum Field Theory} (McGraw-Hill, 1980).

\bibitem{boos}
E.E.Boos, A.I.Davydychev, Theor.Math.Phys. {\bf 89},1,(1991)1052. 

\bibitem{stand}
A.I.Davydychev, {\it Proc. International Conference ''Quarks-92''}
(World Scientific, 1993); hep-ph/9307323.

\bibitem{jmp}
A.I.Davydychev, J.Math.Phys. {\bf 32},4(1991)1052; J.Math.Phys. {\bf
33},1 (1991) 358.

\bibitem{appel}  
P.Appel, J. Kamp\'e de Feriet, {\it Fonctions Hyperg\'eom\'etriques 
et Hypersph\'eriques. Polynomes D'Hermite} (Gauthiers-Villars, Paris 
1926).

\bibitem{thooft}
G.'t Hooft, M.Veltman, Nucl.Phys.{\bf B44} (1972)189; C.G. 
Bollini, J.J. Giambiagi, Nuovo Cim. {\bf B12} (1972)20.

\bibitem{bateman}
A.Erd\'elyi,W.Magnus,F.Oberhettinger and F.G.Tricomi, {\it Higher 
Transcendental Functions} (Mc-Graw-Hill, 1953). 

\bibitem{edinburgo}
A.Erd\'elyi, Proc.Roy.Soc.Edinburgh, {\bf A62}(1948)378.

\bibitem{davyd}
A.I.Davydychev, J.B.Tausk, Nucl.Phys.{\bf B397} (1993)123.

\bibitem{lebedev}
N.N.Lebedev, {\it Special Functions and Applications} (Pretience-Hall,
1965).  E.D.Rainville, {\it Special Functions} (Chelsea Pub.Co., 1960).

\bibitem{letb}
N.I.Ussyukina, A.I.Davydychev, Phys.Lett.{\bf B332}(1994)159.

\bibitem{eden}
R.J.Eden, P.V.Landshoff, D.I.Olive and J.C.Polkinghorne, {\it The
Analytic S-Matrix} (Cambridge Univ.Press, 1966).

\bibitem{feshbach}
P.M.Morse, H.Feshbach, {\it Methods of Theorethical Physics}
(McGraw-Hill, 1953). E.C.Titchmarsh, {\it The Theory of Functions}
(Oxford Univ.Press, 1939). 

\bibitem{london}
R.J.Eden, Proc.Roy.Soc.{\bf A210}(1952)388.

\bibitem{todorov}
I.T.Todorov, {\it Analytic Properties of Feynman Diagrams in Quantum
Field Theory} (Pergamon Press, 1971). 

\bibitem{fairlie}
D.B.Fairlie, P.V.Landshoff, J.Nutall, J.C.Polkinghorne,
J.Math.Phys. {\bf 3}, 4 (1962) 594.

\end{thebibliography}
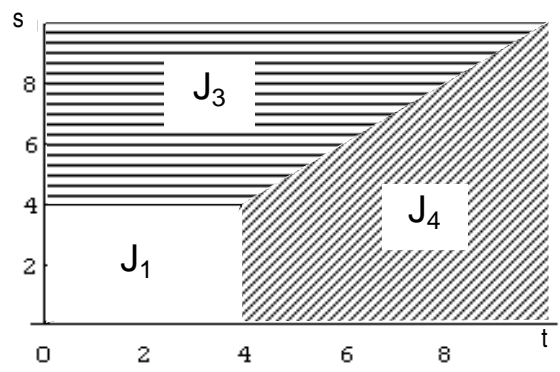
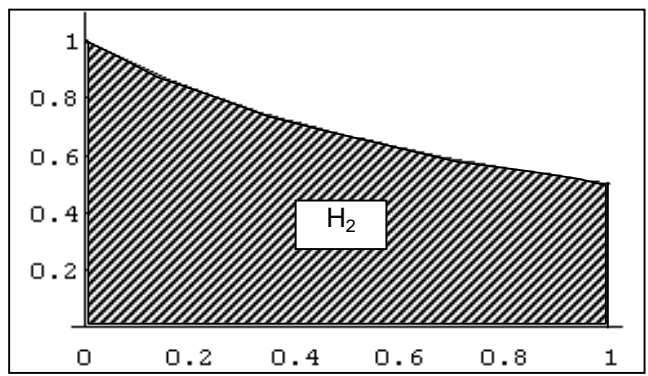
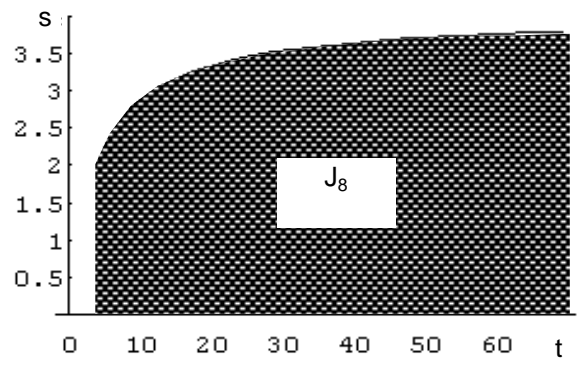
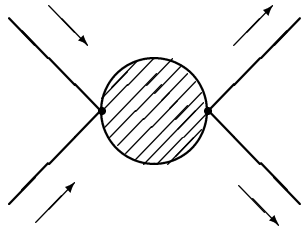
\end{document}